\documentclass[twocolumn,showpacs,preprintnumbers,amsmath,amssymb]{revtex4}
\usepackage{graphicx}

\begin{document}

\title{Fermions in gravity and the skyrmion backgrounds in six dimensional brane-worlds}
\author{Yuta Kodama}
\author{Kento Kokubu}
\author{Nobuyuki Sawado}
\email{sawado@ph.noda.tus.ac.jp}
\affiliation{Department of Physics, Tokyo University of Science, 
Noda, Chiba 278-8510, Japan}

\date{\today}

\begin{abstract}
We construct brane solutions in six dimensional Einstein-Skyrme systems. 
A class of baby skyrmion solutions realize warped compactification of the extra 
dimensions and gravity localization on the brane for negative bulk cosmological constant. 
Coupling of the fermions with the brane skyrmions successfully lead to the 
brane localized fermions. The standard representation of the gamma matrices is used to obtain  
massive localized modes as well as the massless one.
Nonlinear nature of the skyrmions brings richer information 
for the fermions level structure. In terms of the level crossing picture, 
emergence of the massive localized modes as well as the zero mode are observed.  
\end{abstract}

\maketitle 

\section{Introduction}
Theories with extradimensions have been expected to solve the hierarchy 
problem and cosmological constant problem. Experimentally unobserved 
extradimensions indicate that the standard model particles and forces 
are confined to a 3-brane~\cite{arkani-hamed98,randall99-1,randall99-2}. 
Intensive study has been performed for the RS brane model in five space-time 
dimensions~\cite{randall99-1,randall99-2}. In this framework, the exponential 
warp factor in the metric can generate a large hierarchy of scales. 
This model, however, requires unstable negative tension branes  
and the fine-tuning between brane tensions and bulk cosmological constant.
 
There is hope that higher dimensional brane models than five 
could evade those problems appeared in five dimensions. 
In fact brane theories in six dimensions show a very distinct feature towards 
the fine-tuning and negative tension brane problems. 
In Refs.~\cite{carroll03,navarro03}, it was shown that the brane tension merely 
produces deficit angles in the bulk and hence it can take an arbitrary value without 
affecting the brane geometry. The model is based on the spontaneous compactification 
by the bulk magnetic flux. If the compactification manifold is a sphere, two branes 
have to be introduced with equal tensions. If it is a disk, no second 3-brane is 
needed. But still the fine-tuning between magnetic flux and the bulk cosmological constant  
can not be avoided although non-static solutions could be free of any fine-tuning~\cite{kanti01}.  

Alternatively to the flux compactification in 6 dim., the nonlinear sigma model has been 
used for compactifications of the extra space 
dimensions~\cite{gell-mann84,kehagias04,rubakov04,lee05}. 
As in the flux compactification, no second 3-brane is needed if the parameters 
in the sigma model and bulk space-time are tuned.   

Warped compactifications are also possible in six space-time dimensions 
in the model of topological objects such as defects and solitons.  
In this context strings~\cite{cohen99,gregory00,gherghetta00,giovannini01,ringeval05} 
were investigated, showing that they can realize localization of gravity.  
Interestingly, if the brane is modeled in such a field theory language, 
the fine-tuning between bulk and brane parameters required in the case 
of delta-like branes turns to a tuning of the model parameters~\cite{ringeval02}. 

The Skyrme model is known to possess soliton solutions called baby skyrmions 
in two dimensional space~\cite{piette94,piette95}. In this paper we therefore consider 
the warped compactification of the two dimensional extra space by the baby skyrmions.  
We find that in the 6 dim. Einstein-Skyrme systems, static solutions which realize warped 
compactification exist for negative bulk cosmological constant. Since the solution 
is regular except at the conical singularity, it has only single 3-brane. 
Thus no fine-tuning between brane tensions is required. The Skyrme model possess a 
rich class of stable multi soliton solutions. We find various brane solutions by such 
multi-solitons.
 
It should be noted that general considerations in the 6 dim. brane model with bulk scalar fields 
suggest that the mechanism of regular warped compactification with single 
positive tension brane is not possible~\cite{chen00}. However, the model under consideration 
is restricted to the bulk scalar field depending only on the radial coordinate in the extra 
space. The scalar field in the Skyrme model depend not only the radial coordinate but 
also the angular coordinate to exhibit nontrivial topological structure, which 
makes possible to realize regular warped compactification.  

Study of localization of fermions and gauge fields on topological defects have been extensively 
studied with co-dimension one \cite{kahagias01,melfo06,ringeval02f,koley08} and two 
\cite{randjbar-daemi00,libanov01,neronov02,zhao07,randjbar-daemi03,parameswaran07}. 
In many years before, particle localization on a domain wall in higher dimensional
space time was discussed\cite{rubakov83,akama}. 
The authors suggested the possibility of localized massless fermions 
on the 1 dim. kink background in 4+1 space-time with Yukawa-type coupling manner.
Localization of chiral fermions on RS scenario is in Ref.\cite{kahagias01}. 
Analysis for the massive fermionic modes was done by Ringeval {\it et,al.} in Ref.\cite{ringeval02f}.
For co-dimension two, the localization on higher dimensional generalizations of the RS model
was studied by the coupling of real scalar fields \cite{randjbar-daemi00}.
Many studies have been followed and most of them are based on the 
Abelian Higgs model coupled with the chiral fermions. 

Problem of fermion mass hierarchy was discussed in Ref.\cite{arkani-hamed00,dvali00,libanov01,neronov02}
within different mechanisms. Especially, in Ref.\cite{libanov01} the hierarchy between 
the fermionic generations are explained in terms of multi-winding number solutions of 
the complex scalar (Higgs) fields. They observed three chiral fermionic zero modes on a 
topological defect with winding number three and finite masses appear the mixing of those
zero modes. Although any brane localization mechanism is absent in their discussion, 
the idea is promising. 

In this article, we employ somewhat different set up:
we consider the localization of the fermions on the skyrmion branes and,  
to treat the massive fermionic modes directly, we use the standard 
representation of the higher dimensional gamma matrices instead of the chiral one. 
The fermion localization and the existence of the zero modes
are confirmed through the analysis of spectral flow of the one particle state. 

This paper is organized as follows. In the next section we describe the 
Einstein-Skyrme system in six dimensions and derives the coupled equations for 
the Skyrme and gravitational fields. We derive a class of multi-winding number 
solutions. We will show some typical numerical 
solutions. Formulation of the fermions in higher dimensional 
curved space time is discussed in Sec.III. Coupling of the fermions and
the skyrmions is introduced in this section.  
Conclusion and discussion are given in Sec.IV.

\section{Construction of the baby-skyrmion branes}
\subsection{\label{subsec:level11}Model}
We consider the model of the 6 dim. Einstein-Skyrme system with a bulk cosmological constant 
coupled to fermions. The action comprises 
\begin{eqnarray}
	S=S_{\rm gravity}+S_{\rm brane}+S_{\rm fermion}\label{action}\,.
\end{eqnarray}
Here $S_{\rm gravity}$ is the six dimensional Einstein-Hilbert action
\begin{eqnarray} 
	S_{\rm gravity}&=&\int d^{6}x \sqrt{-g}\left[\frac{1}{2\kappa^2}R-\Lambda_{b}\right] 
	\label{grav_action} 
\end{eqnarray}
In the parameter $\kappa^2=1/M_{6}^{4}$, $M_{6}$ is the six-dimensional 
Planck mass, denoted the fundamental gravity scale, and 
$\Lambda_{b}$ is the bulk cosmological constant.

For $S_{\rm brane}$ we use the action of baby-Skyrme model~\cite{piette94,piette95}
\begin{eqnarray}
S_{\rm brane}&=&\int d^{6}x {\cal L}_{\rm brane}
\end{eqnarray}
with
\begin{eqnarray}
	&&{\cal L}_{\rm brane}=\sqrt{-g}\Bigl[\frac{F^2}{2}\partial_{M}
	{\vec \phi}\cdot \partial^{M}{\vec \phi}
	+\frac{1}{4e^2}\bigl(\partial_{M}{\vec \phi}\times\partial_{N}
	{\vec \phi}\bigr)^{2} \nonumber \\
	&&\hspace{5cm}+\mu^2 (1+{\vec n}\cdot{\vec \phi})\Bigr]\,,
	\label{skyrme_action}
\end{eqnarray}
where $M,N$ run over $0,\cdots ,5$ and  ${\vec n}=(0,0,1)$.
 ${\vec \phi}=(\phi^{1},\phi^{2},\phi^{3})$ denotes a triplet of scalar real fields with 
the constraint ${\vec \phi}\cdot{\vec \phi}=1$.
The $F,e,\mu$ are the Skyrme model parameters wih the dimension of $M^{2}$, $M^{-1}$, $M^{3}$ 
respectively. The first term in (\ref{skyrme_action}) is familiar from $\sigma-$model.   
The second term is the analogue of the Skyrme fourth order term in the usual Skyrme model
which works as a stabilizer for obtaining the soliton solution. The last term is referred to
as a potential term which guarantee the stability of a baby-skyrmion. 

The solutions of the 
model would be characterized by following topological charge in curved space-time
\begin{eqnarray}
Q=\frac{1}{4\pi}\int d^2x{\vec \phi}\cdot (\nabla_1{\vec \phi}\times \nabla_2{\vec \phi})
\label{windingnumber}
\end{eqnarray}
where $\nabla_\mu$ means the space-time covariant derivative. 
Let us assume that the matter Skyrme fields depend only on the 
extra coordinates and impose the hedgehog ansatz 
\begin{eqnarray}
      {\vec \phi}=(\sin f(r)\cos n\theta,\, \sin f(r)\sin n\theta,\, \cos f(r))\,. 
      \label{hedgehog}
\end{eqnarray}
The function $f(r)$ which is often called profile function, has following boundary condition
\begin{eqnarray}
f(0)=-(m-1)\pi,~~\lim_{r\to \infty}f(r)=\pi
\end{eqnarray}
where $(m,n)$ is arbitrary integer. This ansatz ensures the topological charge
\begin{eqnarray}
Q=n(1-(-1)^m)/2.
\label{windingnumber2}
\end{eqnarray}

We consider the maximally symmetric metric with vanishing 4D cosmological constant, 
\begin{eqnarray}
	ds^{2}=B^{2}(r)\eta_{\mu\nu}dx^{\mu}dx^{\nu} + dr^{2}
	+ C^{2}(r)d\theta^{2} \label{metric}
\end{eqnarray}
where $\eta_{\mu\nu}$ is the Minkowski metric with the signature 
$(-,+,+,+)$ in our convention and $0\le r < \infty$ and
 $0\le \theta \le 2\pi$. 
This ansatz has been proved to realize warped compactification 
of the extra dimension in models where branes are represented 
by global defects~\cite{olasagasti00}. 

$S_{\rm fermion}$ is the action of a fermions coupled with the skyrmions and the warp factors; that 
would be described in Sec.\ref{sec:level3}.

The general forms of the coupled system of Einstein equations and the 
equation of motion of the Skyrme model are 
\begin{eqnarray}
&&G_{MN}=\kappa^2 (-\Lambda_{b} g_{MN} +T_{MN})\,, \\
&&\frac{1}{\sqrt{-g}}\partial_N\Bigl(\sqrt{-g}F^2{\vec \phi}\times \partial^N{\vec \phi} \nonumber \\
&&+\sqrt{-g}\frac{1}{e^2}\partial_M{\vec \phi}\bigl(\partial^M{\vec \phi}\cdot({\vec \phi}\times \partial^N{\vec \phi})\bigr)\Bigr)
+\mu^2{\vec \phi}\times{\vec n}=0\,,\nonumber \\
\end{eqnarray}
where the stress-energy tensor $T_{MN}$ is given by
\begin{eqnarray}
&&T_{MN}=-2\frac{\delta {\cal L}_{\rm brane}}{\delta g^{MN}}+g_{MN}{\cal L}_{\rm brane} \nonumber \\
&&=F^2\partial_{M}{\vec \phi}\cdot \partial_{N}{\vec \phi}
	+\frac{1}{e^2}g^{AB}\bigl(\partial_{A}{\vec \phi}\times\partial_{M}{\vec \phi}\bigr)
	\cdot\bigl(\partial_{B}{\vec \phi}\times\partial_{N}{\vec \phi}\bigr)\nonumber \\
	&&+g_{MN}{\cal L}_{\rm brane}\,.
	\label{stress_tensor}
\end{eqnarray}
Inserting Eq.~(\ref{hedgehog}) into Eq.~(\ref{skyrme_action}), one obtains 
the Lagrangian 
\begin{eqnarray}
	{\cal L}_{\rm brane}=-B^4{\tilde C}F^4e^2\left[uf'^{2}+\frac{n^2\sin^{2}f}{{\tilde C^{2}}}
	+2{\tilde \mu}(1+\cos f)\right] \nonumber \\
	\label{skyrme-lag}
\end{eqnarray}
where we have introduced the dimensionless quantities 
\begin{eqnarray}
	\tilde{x}_\mu=eFx_\mu,~y=eFr,~{\tilde C}=eFC,~
	{\tilde \mu}=\frac{1}{eF^2}\mu \label{}
\end{eqnarray}
and 
\begin{eqnarray}
	u=1+\frac{n^2\sin^{2}f}{{\tilde C}^{2}}\,. \label{}
\end{eqnarray}
The prime denotes derivative with respect to the radial component $y$ of the two extra space. 
The Skyrme field equation is thus
\begin{eqnarray}
	&&f''+\left(\frac{4B'}{B}+\frac{{\tilde C}'}{{\tilde C}}+\frac{u'}{u}
	\right)f' \nonumber \\
	&&-\frac{1}{2u}\left[\frac{n^2\sin 2f}{{\tilde C}^{2}}
	(1+f'^{2})+2{\tilde \mu}^2\sin f\right]=0 \label{skyrme}
\end{eqnarray}
where 
\begin{eqnarray}
	\frac{u'}{u}=\frac{n^2}{{\tilde C}^{2}+n^2\sin^{2}f}\left[f'\sin 2f
	-2\frac{{\tilde C}'}{{\tilde C}}\sin^{2}f\right]\,. \label{}
\end{eqnarray}
Within this ansatz, the components of the stress-energy tensor
 (\ref{stress_tensor}) becomes
\begin{eqnarray}
&&T_{\mu\nu}=-F^4e^2\tilde{B}^2\eta_{\mu\nu}\tau_0(y)\,,~~\nonumber \\
&&~~~~\tau_0(y)\equiv \frac{u}{2}f'^2+\frac{n^2\sin^2f}{2\tilde{C}^2}+\tilde{\mu}^2(1+\cos f)
\label{estensor0}\\
&&T_{rr}=-F^4e^2\tau_r(y)\,,~~\nonumber \\
&&~~~~\tau_r(y)\equiv -\frac{u}{2}f'^2+\frac{n^2\sin^2f}{2\tilde{C}^2}+\tilde{\mu}^2(1+\cos f)
\label{estensorr}\\
&&T_{\theta\theta}=-F^4\tilde{C}^2\tau_\theta(y)\,,\nonumber \\
&&~~~~\tau_\theta(y)\equiv \frac{\hat{u}}{2}f'^2-\frac{n^2\sin^2f}{2\hat{C}^2}+\tilde{\mu}^2(1+\cos f)
\label{estensort}
\end{eqnarray}
where 
\begin{eqnarray}
	\hat{u}=1-\frac{n^2\sin^{2}f}{{\tilde C}^{2}}\,. \label{}
\end{eqnarray}
The Einstein equations with bulk cosmological constant are written 
down in the following form
\begin{eqnarray}
	&& 3{\hat b}'+6{\hat b}^{2}+3{\hat b}\,{\hat c}+{\hat c}'+{\hat c}^2
	=-\alpha({\tilde \Lambda}_{b}+\tau_0(y)) \label{einstein1}\\ 
	&& 6{\hat b}^{2}+4{\hat b}\,{\hat c}=-\alpha
	({\tilde \Lambda}_{b}+\tau_r(y)) \label{einstein2}\\
	&& 4{\hat b}'+10{\hat b}^{2}=-\alpha({\tilde \Lambda}_{b}
	+\tau_\theta(y))\label{einstein3}
\end{eqnarray}
where $\alpha=\kappa^2 F^2$ is a dimensionless coupling constant and 
${\tilde \Lambda}_{b}=\Lambda_{b} /e^2F^{4}$ is a dimensionless bulk cosmological constant. 
Also, we introduce ${\hat b}\equiv B'/B,{\hat c}\equiv C'/C$ for convenience. 

\subsection{Boundary conditions}
At infinity, all components of the energy-momentum tensor vanishes and the Einstein 
equations~(\ref{einstein1})-(\ref{einstein3}) are then reduced to 
\begin{eqnarray}
	&&3{\hat b}'+6{\hat b}^{2}+3{\hat b}\,{\hat c}+{\hat c}'+{\hat c}^2
	=-\alpha {\tilde \Lambda}_{b} \\
	&&6{\hat b}^{2}+4{\hat b}\,{\hat c}=-\alpha 
	{\tilde \Lambda}_{b} \\
	&&4{\hat b}'+10{\hat b}^{2}=-\alpha {\tilde \Lambda}_{b}\,.\label{}
\end{eqnarray}
The general solution has been obtained in Ref.~\cite{giovannini01,ringeval05} which 
is given by
\begin{eqnarray}
	{\hat b}=p\,\frac{Ae^{\frac{5}{2}py}-e^{-\frac{5}{2}py}}
	{Ae^{\frac{5}{2}py}+e^{-\frac{5}{2}py}}\;,\;\;\;\;\;
	{\hat c}=\frac{5p^{2}}{2{\hat b}}
	-\frac{3}{2}{\hat b} \label{}
\end{eqnarray}
where $A$ is an arbitrary constant and 
\begin{eqnarray}
	p=\sqrt{\frac{-\alpha {\tilde \Lambda}_{b}}{10}}\,. \label{}
\end{eqnarray}
Since we are interested in regular solutions with warped compactification of the 
extra-space, the functions $B$ and ${\tilde C}$ must converge at infinity. 
This is achieved only when ${\tilde \Lambda}_{b}<0$ and $A=0$ with the solution
\begin{eqnarray}
	B \rightarrow \epsilon_{1}\,e^{-px}
	\;,\;\;\;\;
	{\tilde C} \rightarrow \epsilon_{2}\,e^{-px} \label{bound-inf1}
\end{eqnarray}
where $\epsilon_{1}$ and $\epsilon_{2}$ are arbitrary constants.
Then, the asymptotic form of the metric which realizes warped compactification 
is given by 
\begin{eqnarray}
	ds_{\infty}^{2}&=&\epsilon_{1}e^{-2\sqrt{\frac{-\alpha {\tilde \Lambda}_{b}}{10}}y}
	\eta_{\mu\nu}dx^{\mu}dx^{\nu} \nonumber \\
	&+&dy^{2}+\epsilon_{2}e^{-2\sqrt{\frac{-\alpha {\tilde \Lambda}_{b}}{10}}y}
	d\theta^{2}\,. 
	\label{asym-met}
\end{eqnarray}

The four-dimensional reduced Planck mass $M_{pl}$ is derived by the 
coefficient of the four-dimensional Ricci scalar, which can be calculated 
inserting the metric~(\ref{metric}) into the action~(\ref{grav_action}), 
\begin{eqnarray*}
	&&\frac{M_{pl}^{2}}{2}\int d^{4}x \sqrt{-g^{(4)}}R^{(4)}
	=\frac{M_{6}^{4}}{2}\int d^{6}x \sqrt{-g}B^{-2}(r)R^{(4)} \\
	&&=\frac{M_{6}^{4}}{2}\int d^{4}x \sqrt{-g^{(4)}}R^{(4)}
	\int dr d\theta B^{2}(r)C(r) \\
	&&= \frac{2\pi M_{6}^{4}}{2}\int dr \, B^{2}(r)C(r)
	\int d^{4}x \sqrt{-g^{(4)}}R^{(4)}  \label{}
\end{eqnarray*}
where the superscript $(4)$ represents a tensor defined on the four-dimensional 
submanifold.  
Thus, we find the relation between $M_{pl}$ and $M_{6}$ as 
\begin{eqnarray}
	M_{pl}^{2}=2\pi M_{6}^{4}\int_{0}^{\infty}dr \,B^{2}(r) C(r) \,. 
\end{eqnarray}
The requirement of gravity localization is equivalent to the finiteness 
of the four-dimensional Planck mass. For the asymptotic 
solution~(\ref{asym-met}), the localization is attained. 

\begin{figure}
\includegraphics[height=7.0cm, width=9cm]{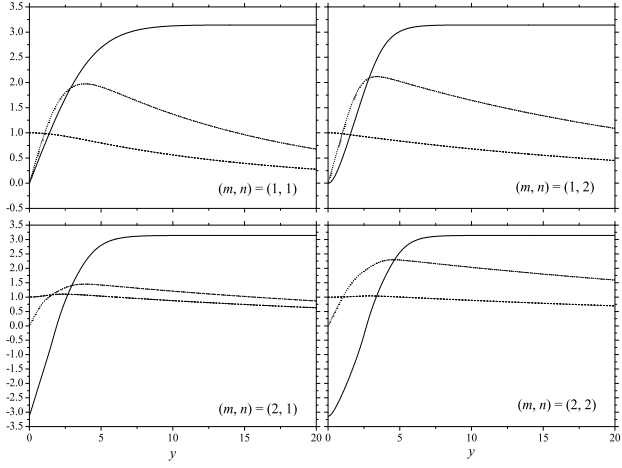}
\caption{\label{fig1} Typical results of the profile functions $f$ (straight line), 
the warp metrices $B,\tilde{C}$ (dashed,dotted line,respectively), 
as a function of $y$. }
\end{figure}

Let us consider the asymptotic solutions for skyrmions.
we can write 
\begin{eqnarray}
f(y)=\bar{f}+\delta f(y)\,,
\end{eqnarray} 
where for $y\gg 1$, $\bar{f}\sim 0$.
The linearized field equations are given by
\begin{eqnarray}
	\delta f''-5p\delta f' -{\tilde \mu}\delta f=0\,. \label{}
\end{eqnarray} 
Assuming that $f$ falls off exponentially, one obtains for $x\gg 1$
\begin{eqnarray}
	\delta f(y) \rightarrow f_{c}e^{-qy}~~~~{\rm with}~~~~ 
	q=\frac{\sqrt{25p^{2}+4{\tilde \mu}}-5p}{2}\;\;\; 
	\label{bound-inf2}
\end{eqnarray}
where $f_{c}$ is an arbitrary constant. 

Following regularity of the geometry at the center of the defect
are imposed  
\begin{eqnarray}
B'(0)=0,~~C(0)=0,~~C'(0)=1
\end{eqnarray}
and we can arbitarily fix $B(0)=1$.
Boundary conditions for the warp factors and the profile function at the origin are 
determined by expanding them around the origin. For the different topological sectors, 
the first few terms are schematically written down as  
\begin{eqnarray}
&&f(y)=-(m-1)\pi+f^{(n)}(0)y^n+O(y^{n+1}) \\
&&b(y)={\cal B}y+O(y^3) \\
&&\tilde{C}(y)=y+{\cal C}y^3+O(y^5)
\end{eqnarray}
where
\begin{eqnarray}
&&(m,n)=(1,1) \nonumber \\
&&~~~~~~{\cal B}=-\frac{\alpha}{4}\Bigl(\tilde{\Lambda}_b+2\tilde{\mu}-\frac{1}{2}f'(0)^4\Bigr)\,,\nonumber \\
&&~~~~~~{\cal C}=\frac{\alpha}{12}\Bigl(\tilde{\Lambda}_b+2\tilde{\mu}-2f'(0)^2 
-\frac{5}{2}f'(0)^4\Bigr) \\
&&(m,n)=(1,2) \nonumber \\
&&~~~~~~{\cal B}=-\frac{\alpha}{4}\bigl(\tilde{\Lambda}_b+2\tilde{\mu}),~~
{\cal C}=\frac{\alpha}{12}\bigl(\tilde{\Lambda}_b+2\tilde{\mu}) \\
&&(m,n)=(2,1) \nonumber \\
&&~~~~~~{\cal B}=-\frac{\alpha}{4}\Bigl(\tilde{\Lambda}_b-\frac{1}{2}f'(0)^4\Bigr),\nonumber \\
&&~~~~~~{\cal C}=\frac{\alpha}{12}\Bigl(\tilde{\Lambda}_b
-2f'(0)^2-\frac{5}{2}f'(0)^4\Bigr) \\
&&(m,n)=(2,2) \nonumber \\
&&~~~~~~{\cal B}=-\frac{\alpha}{4}\tilde{\Lambda}_b,~~
{\cal C}=\frac{\alpha}{12}\tilde{\Lambda}_b\,.
\end{eqnarray}
Thus one finds that the only $f'(0)$ or $f''(0)$ is the free parameter vicinity of the orgin.

\begin{figure}
\includegraphics[height=7.0cm, width=9cm]{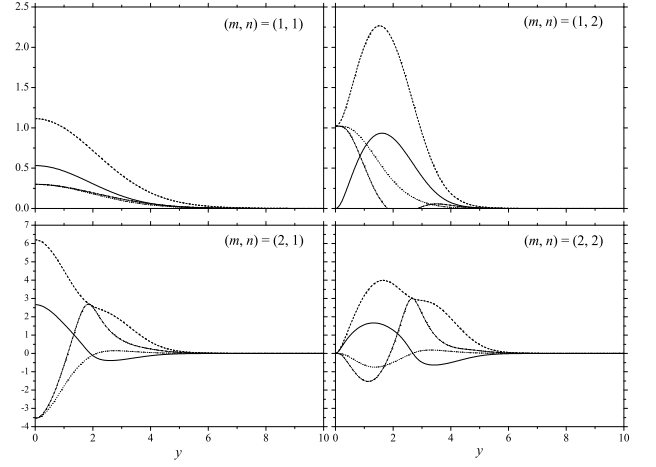}
\caption{\label{fig2} Typical results of the stress-energy tensors $\tau_0,\tau_r,\tau_\theta$ 
(dashed,dotted,and dot-dashed line, respectively)and the topological charge
density $q$ (straight line).}
\end{figure}

Consider linear combinations of Eqs.(\ref{einstein1})-(\ref{einstein3}), we obtain
\begin{eqnarray}
&&{\hat b}'+4{\hat b}^2+{\hat b}\,{\hat c}=
-\frac{1}{2}\alpha \tilde{\Lambda}_b
+\frac{\alpha}{4}(\tau_r+\tau_\theta)\,, \label{einstein5} \\
&&4{\hat b}\,{\hat c}+{\hat c}'+{\hat c}^2=
-\frac{1}{2}\alpha \tilde{\Lambda}_b+\frac{\alpha}{4}(4\tau_0+\tau_r-3\tau_\theta)\,. \label{einstein6}
\end{eqnarray}
Integrating Eqs.(\ref{einstein5}),(\ref{einstein6}) from zero to $x_c$, we get 
\begin{eqnarray}
&&B^3(x_c)B'(x_c)\tilde{C}(x_c) \nonumber \\
&&=-\frac{\alpha}{2}\tilde{\Lambda}_b\int^{x_c}_0 B^4 \tilde{C} dx
-\frac{\alpha}{4}(\mu_r+\mu_\theta)\,,
\label{einstein7}\\
&&B^4(x_c)\tilde{C}'(x_c) \nonumber \\
&&=1-\frac{\alpha}{2}\tilde{\Lambda}_b\int^{x_c}_0 B^4 \tilde{C} dx
-\frac{\alpha}{4}(4\mu_0+\mu_r-\mu_\theta)\,.
\label{einstein8} 
\end{eqnarray}
(\ref{einstein7}) is the six-dimensional analogue of the relation determining the 
Tolman mass whereas Eq.(\ref{einstein8}) is the generalization of the relation giving the 
angular deficit. 
Combining these the following relations are obtained in the $x_c\to \infty$
\begin{eqnarray}
\alpha\int^{\infty}_0 B^4 \frac{n^2\sin^2 f}{\tilde{C}}(1+f'^2)dx=1 
\end{eqnarray}
or
\begin{eqnarray}
\alpha\int^{\infty}_0 B^4\Bigl[\frac{n^2\sin^2 f}{\tilde{C}}+2\tilde{\Lambda}_b\tilde{C}
+2\tilde{\mu}\tilde{C}(1-\cos f)\Bigr]dx=1.\nonumber \\
\end{eqnarray}
These conditions are used for checking the numerical accuracy of our calculations.  

\subsection{Numerical solutions}
The equations~(\ref{skyrme}),(\ref{einstein1})-(\ref{einstein3}) should be solved numerically since 
they are highly nonlinear. The simple technique to solve the Einstein-Skyrme 
equations are the shooting method combined with the 4th order Runge-Kutta forward 
integration~\cite{shiiki05}. However, a unique set of boundary conditions at $x=0$ 
produces 2 distinct solutions, one of which grows exponentially and the other 
decays exponentially as $x\rightarrow \infty$. This causes instability of 
solutions when the forward integration is performed. Instead, we use a 
backward integration following Refs.~\cite{giovannini01} where the 6 dim. 
vortex-like regular brane solutions were constructed. The backward integration 
method requires a set of boundary conditions at infinity. 
We, however, truncate and take the distance $x_{\rm max}$ far enough from the origin so that 
the Skyrme profile would fall off before it reaches $x_{\rm max}$. 
The set of boundary conditions at $x=x_{\rm max}$ produces a unique solution which 
satisfies the boundary conditions at $x=0$ and hence it is numerically stable.
We present our typical numerical results in Figs.[\ref{fig1},\ref{fig2}].

\section{\label{sec:level3}Fermions}
\subsection{\label{subsec:level31}Basic formalism}
The action of the fermions coupled with the Skyrme field in a Yukawa coupling manner 
can be written as
\begin{eqnarray}
S_{\rm fermion}=\int d^6x {\cal L}_{\rm fermion}
\end{eqnarray}
with
\begin{eqnarray}
{\cal L}_{\rm fermion}=\sqrt{-g}\Bigl[\bar{\Psi}(i\Gamma^AD_A-M\vec{\tau}\cdot\vec{\phi})\Psi\Bigr]\,.
\end{eqnarray}
The six-dimensional gamma matrices $\Gamma^A$ are defined with the help the {\it vielbein} $e^A_{\hat{a}}$
and those of the flat-space $\gamma^{\hat{a}}$,{\it i.e.}, $\Gamma^A:=e^A_{\hat{a}}\gamma^{\hat{a}}$.
The covariant derivative is defined as 
\begin{eqnarray} 
D_A:=\frac{1}{2}\overleftrightarrow{\partial}_A+\frac{1}{2}\omega_A^{\hat{a}\hat{b}}\sigma_{\hat{a}\hat{b}}
\end{eqnarray}
where $\omega_A^{\hat{a}\hat{b}}:=\frac{1}{2}e^{\hat{a}B}\nabla_Ae^{\hat{b}}_B$ are the spin connection 
with generators $\sigma_{\hat{a}\hat{b}}:=\frac{1}{4}[\gamma_{\hat{a}},\gamma_{\hat{b}}]$.
The simbol $\overleftrightarrow{\partial}$ implies that $\psi\overleftrightarrow{\partial}\phi\equiv \psi\partial \phi-(\partial \psi) \phi$.
Here $A,B=0,\cdots,5$ are the six-dimensional space-time index and $\hat{a},\hat{b}=0,\cdots,5$ corresponds to the flat
tangent six-dimensional Minkowski space. 
The vielbein is defined through $g_{AB}=e^{\hat{a}}_Ae_{\hat{a}B}=\eta_{\hat{a}\hat{b}}e^{\hat{a}}_Ae^{\hat{b}}_B$.
The definition which was introduced in Refs.\cite{neronov02,randjbar-daemi03} is simply defined as
\begin{eqnarray}
&&e^{\hat{a}}_\mu=B(r) \delta^{\hat{a}}_\mu,~~\mu=0,\cdots,3,\nonumber \\
&&e^{\hat{r}}_r=1,~~e^{\hat{\theta}}_\theta=C(r).
\end{eqnarray}
They are the definitions which produce the gamma matrices parallel to the polar 
coordinate $(r,\theta)$. Thus, special care is needed to explore 
the conserved quantities like angular momenta from the corresponding hamiltonian.
Non-vanishing components of the corresponding spin connection are found to be
\begin{eqnarray}
&&\omega^{\hat{\mu}\hat{4}}_\mu=\delta^{\hat{\mu}}_{\mu}B'(r),~~
\omega^{\hat{4}\hat{5}}_5=-C'(r)
\end{eqnarray}
where $'\equiv \frac{d}{dr}$.
The Dirac equation takes the form
\begin{eqnarray}
&&\Bigl[i\frac{1}{B}\delta^{\mu}_{\hat{\mu}}\gamma^{\hat{\mu}}\partial_\mu
+i\gamma^{\hat{4}}
(\partial_r+\frac{2B'}{B}+\frac{C'}{2C}) \nonumber \\
&&+i\gamma^{\hat{5}}\frac{1}{C}\partial_\theta
-M\vec{\tau}\cdot\vec{\phi}\Bigr]\Psi=0\,.
\label{diraceq8_s_0}
\end{eqnarray}
The corresponding hamiltonian reads
\begin{eqnarray}
H=-i\gamma^{\hat{0}}\gamma^{\hat{4}}(\partial_r+\frac{2B'}{B}+\frac{C'}{2C})
-i\gamma^{\hat{0}}\gamma^{\hat{5}}\frac{1}{C}\partial_\theta +\gamma^{\hat{0}}M\vec{\tau}\cdot\vec{\phi}\,.
\nonumber \\
\label{hamiltonian8_old}
\end{eqnarray}
The equation (\ref{diraceq8_s_0}) can be solved numerically in terms of, for example, the shooting method. 
At present study, however, we will treat the problem by somewhat different way.
According to the Rayleigh-Ritz variational method \cite{bransden}, the upper 
bound of the spectrum can be obtained from the secular equation; 
\begin{eqnarray}
	\rm{det}\left(\bf{A}-\epsilon \bf{B}\right) = 0 
	\label{secular_equation}
\end{eqnarray}
where 
\begin{eqnarray*}
     A_{ij}=
	\int d^{3}x \varphi_i^\dagger H
	\varphi_j,
	~~B_{ij}=
	\int d^{3}x \varphi_i^\dagger \varphi_j
\end{eqnarray*} 
where $\{\varphi_i\}~(i=1,\cdots,N)$ is some complete set of the plane-wave spinor basis. 
For $N \rightarrow \infty$, the spectrum $\epsilon$ becomes exact. 
Eq.(\ref{secular_equation}) can be solved numerically. 
For simplicity, we are to construct plane-wave basis in the flat space-time, {\it i.e.},$B=1,B'=0,C=r$,
as $\{\varphi_i\}$. However, no corresponding flat Hamiltonian subject to Eq.(\ref{hamiltonian8_old}) 
exists; thus it is not easy task to construct the plane-wave basis. 

\begin{figure}
\includegraphics[height=12cm, width=9cm]{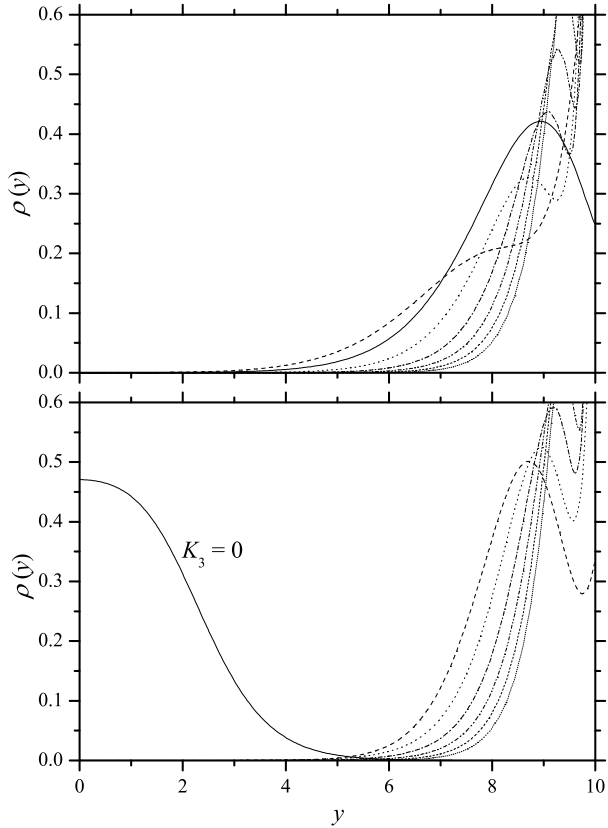}
\caption{\label{fig3} Fermion number density for the background skyrmion with $(m,n)=(1,1)$.
The results for the coupling constant $m=0$ (decoupled) and $m=0.86$ are plotted. 
If fermions couple to the skyrmions, only the state $K_3=0$ is localized on the brane core.
The all other states are not observed because they are strongly delocalized.}  
\end{figure}

\begin{figure}
\includegraphics[height=7.0cm, width=9cm]{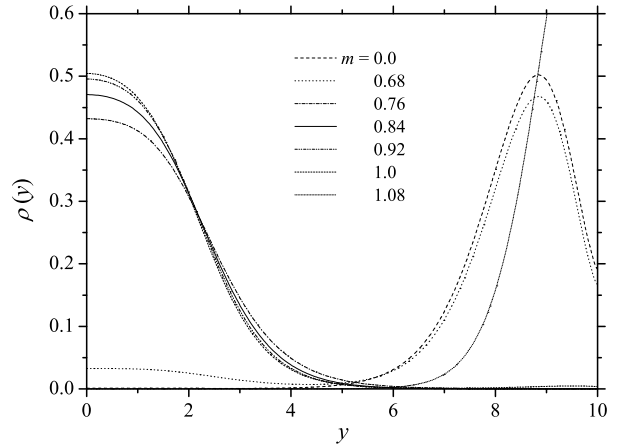}
\caption{\label{fig4} Fermion number density for the background skyrmion with $(m,n)=(1,1)$
for the varying the coupling constants.}
\end{figure}

\begin{figure}
\includegraphics[height=7.0cm, width=9cm]{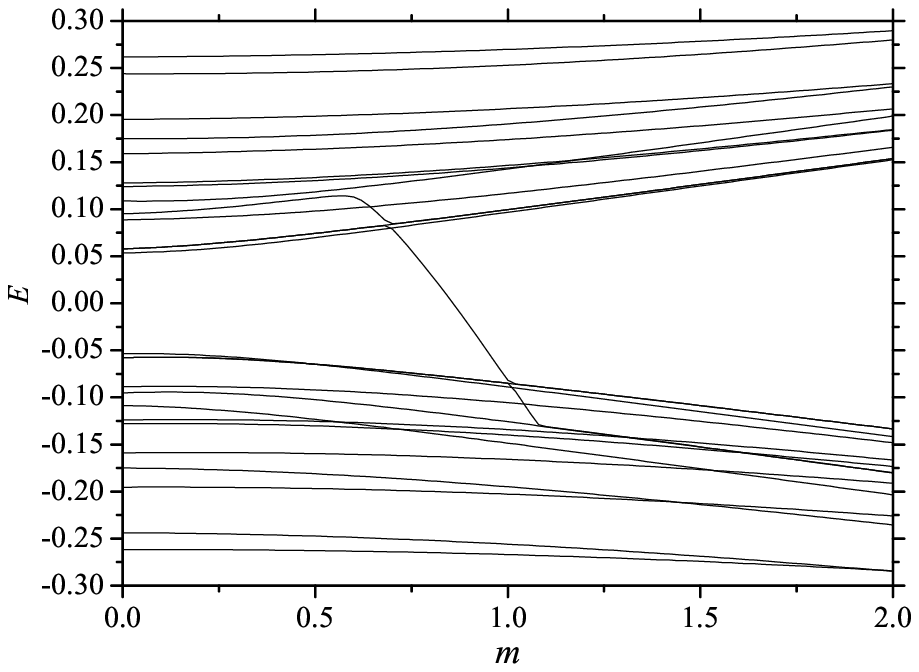}
\caption{\label{fig5} Spectral flow of the fermion energy of $(m,n)=(1,1)$, with some excited energy spectra.
The zero crossing spectrum has the quantum number $K_3=0$, which corresponds to the localizing zero 
mode.}
\end{figure}

\begin{figure*}
\includegraphics[height=5.0cm, width=6cm]{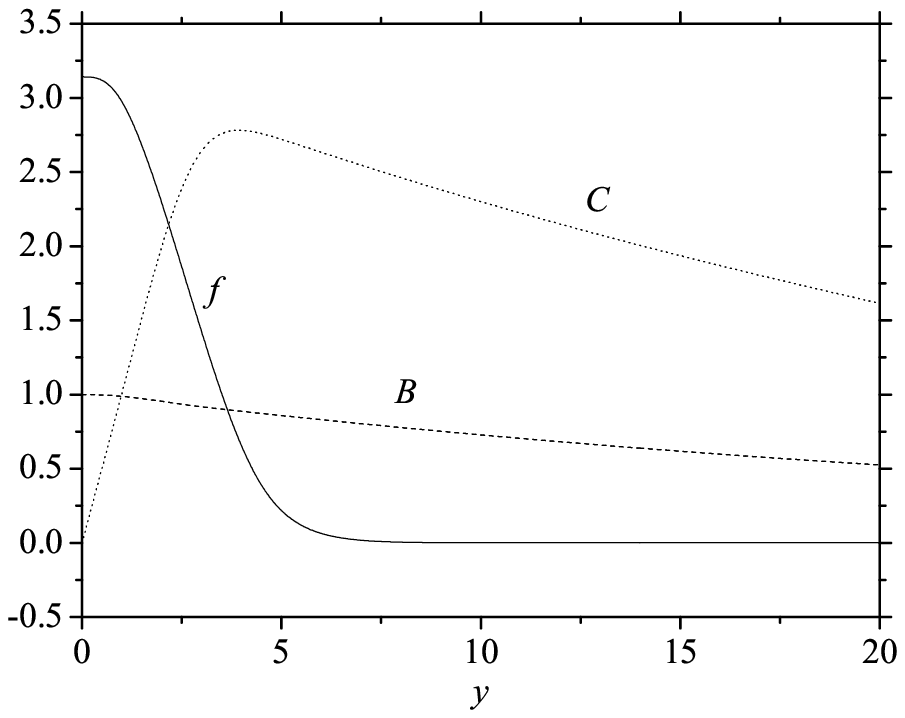}
\includegraphics[height=5.0cm, width=6cm]{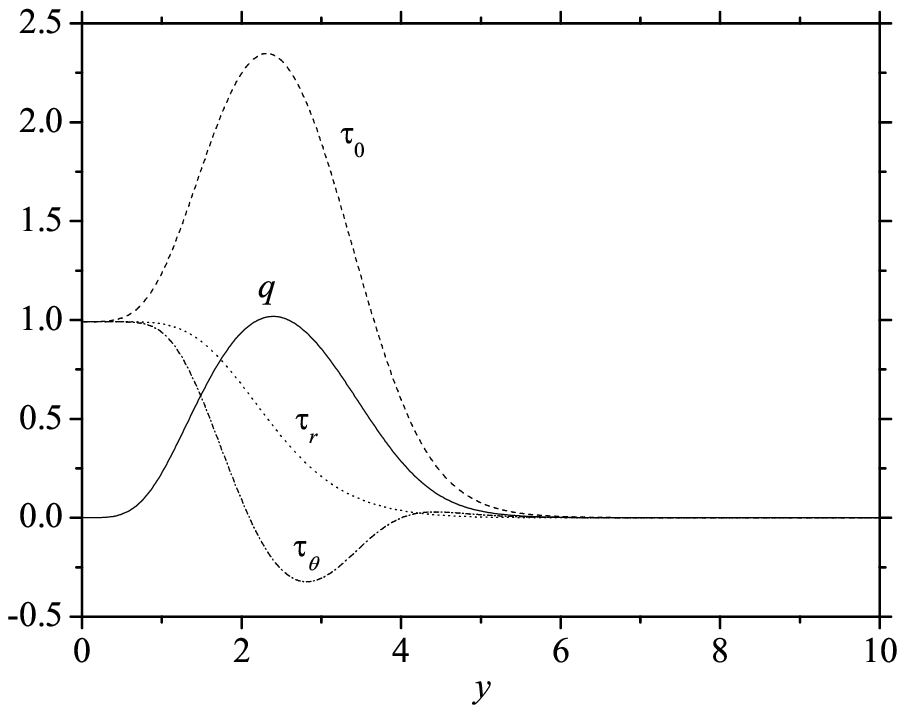}
\caption{\label{fig6} The typical brane solution with $(m,n)=(1,3)$; the profile function and the warp factors (left), 
and the stress energy tensors and the topological charge (right).}
\end{figure*}

We employ a different form of the vielbein which was used at, {\it e.g.}, Ref.\cite{zhao07}, that is
\begin{eqnarray}
&&e^{\hat{a}}_\mu=B(r) \delta^{\hat{a}}_\mu,~~\mu=0,\cdots,3, \nonumber \\
&&e^{\hat{4}}_r=\cos \theta,~~e^{\hat{5}}_r=\sin \theta,~~\nonumber \\
&&e^{\hat{4}}_\theta=C(r)\sin\theta,~~e^{\hat{5}}_\theta=C(r)\cos\theta\,.
\end{eqnarray}
The new definition produces the gamma matrices parallel to the Cartesian 
coordinate $(x_4,x_5)$. 
Non-vanishing components of the corresponding spin connections are then
\begin{eqnarray}
&&\omega^{\hat{\mu}\hat{4}}_\mu=\delta^{\hat{\mu}}_{\mu}B'(r)\cos\theta,~~
\omega^{\hat{\mu}\hat{5}}_\mu=\delta^{\hat{\mu}}_{\mu}B'(r)\sin\theta,~~\nonumber \\
&&\omega^{\hat{4}\hat{5}}_\theta=1-C'(r).
\end{eqnarray}
The Dirac equation is now 
\begin{eqnarray}
&&\Bigl[i\frac{1}{B}\delta^{\mu}_{\hat{\mu}}\gamma^{\hat{\mu}}\partial_\mu
+i(\cos\theta\gamma^{\hat{4}}+\sin\theta\gamma^{\hat{5}})
(\partial_r+\frac{2B'}{B}-\frac{1-C'}{2C}) \nonumber \\
&&-i(\sin\theta\gamma^{\hat{4}}-\cos\theta\gamma^{\hat{5}})
\frac{1}{C}\partial_\theta
-M\vec{\tau}\cdot\vec{\phi}\Bigr]\Psi=0\,.
\label{diraceq8_s}
\end{eqnarray}
The Dirac gamma matrices should satisfy the anti-commutation relations 
$\{\gamma^{\hat{A}},\gamma^{\hat{B}}\}=2\eta^{\hat{A}\hat{B}}$
and there are the candidates preserving such Clifford algebra. 
In most of previous studies in 6 dim., their analyses are based on the localization 
on the Abelian vortex and the chiral representation of gamma matrices. 
In this representation, the spinor
can be expanded into the right and the left components. The zero mode appears as 
a eigenstate of right or left component. 
For the massive modes, they can be estimated from the mixing of both components. 
To treat the massive fermionic modes directly, 
we employ the standard representation of the higher dimensional gamma matrices 
instead of the chiral one. 
Since the eigenvalues depend on the properties of the background brane solutions, 
zero modes should appear as a special case of the massive modes. 
The standard representation of the gamma matrices in 6 dim. can be defined as
\begin{eqnarray}
&&\gamma^{\hat{\mu}}:=
\left(
\begin{array}{cc}
\bar{\gamma}^{\hat{\mu}}& 0 \\
0 & -\bar{\gamma}^{\hat{\mu}} \\
\end{array}
\right),~
\bar{\gamma}^{\hat{\mu}}\equiv \biggl\{
\left(
\begin{array}{cc}
I_2 & 0 \\
0 & -I_2 \\
\end{array}
\right),
\left(
\begin{array}{cc}
0 & \vec{\sigma} \\
-\vec{\sigma} & 0 \\
\end{array}
\right)\biggr\} \nonumber \\
&&~~~~~~~~\gamma^{\hat{4}}:=
\left(
\begin{array}{cc}
0 & -iI_4 \\
-iI_4 & 0 \\
\end{array}
\right),
\gamma^{\hat{5}}:=
\left(
\begin{array}{cc}
0 & -I_4 \\
I_4 & 0 \\
\end{array}
\right)
\end{eqnarray}
where $I_n$ means the identity matrix of the dimension $n$.
The six dimensional spinor $\Psi$ can be decomposed into the four dimensional and 
the extra space-time components
\begin{eqnarray}
\Psi(x^\mu,r,\theta)=\psi(x^\mu)
\left(
\begin{array}{c}
U(r,\theta) \\
V(r,\theta) \\
\end{array}
\right)\,.
\label{wf8_s}
\end{eqnarray}
Here the four dimensional part $\psi(x^\mu)$ is the solution of the corresponding 
Dirac equation on the brane
\begin{eqnarray}
i\bar{\gamma}^\mu\partial_\mu\psi=w\psi
\label{diraceq4}
\end{eqnarray}
in which the eigenvalues $w$ indicate an 4 dim. effective mass of the fermions. 
Substituting the ansatz (\ref{wf8_s}) and (\ref{diraceq4})
into the Dirac equation (\ref{diraceq8_s_0}) yields the equations for $(U,V)$
\begin{eqnarray}
-Be^{-i\theta}(\partial_y-\frac{i\partial_\theta}{\tilde{C}}+{\cal A}_y)V+
Bm\vec{\tau}\cdot\vec{\phi}U=\tilde{w}U && \nonumber \\
Be^{i\theta}(\partial_y+\frac{i\partial_\theta}{\tilde{C}}+{\cal A}_y)U-
Bm\vec{\tau}\cdot\vec{\phi}V=\tilde{w}V 
\label{diraceq8_s}
\end{eqnarray}
where ${\cal A}_y:=2B'/B-(1-\tilde{C}')/2\tilde{C}$. Here we have introduced
the dimensionless coupling constant $m:=M/eF$ (and $\tilde{\omega}:=\omega/eF$).
In order to eliminate such induced potential ${\cal A}_y$ from the equations,
it is convenient to replace the eigenfunctions $(U,V)$ into $(u,v)$~\cite{randjbar-daemi03}
by
\begin{eqnarray}
&&\left(
\begin{array}{c}
U \\
V \\
\end{array}
\right) \nonumber \\
&&=
\exp\biggl[-2\ln B(y)-\frac{1}{2}\ln \tilde{C}(y)+\frac{1}{2}\int^y \frac{dy'}{\tilde{C}(y')}\biggr]
\left(
\begin{array}{c}
u \\
v \\
\end{array}
\right)\,. \nonumber 
\end{eqnarray}
The eigenproblem \label{diraceq8_s} thus reads
\begin{eqnarray}
&&H
\left(
\begin{array}{c}
u \\
v \\
\end{array}
\right)
=\tilde{w}
\left(
\begin{array}{c}
u \\
v \\
\end{array}
\right) 
\label{eigenproblem8}
\\
&&\hspace{-0.5cm}H:=
B\left(
\begin{array}{cc}
m\vec{\tau}\cdot\vec{\phi} &
-e^{-i\theta}(\partial_y-\frac{i\partial_\theta}{\tilde{C}})
\\
e^{i\theta}(\partial_y+\frac{i\partial_\theta}{\tilde{C}})
&
-m\vec{\tau}\cdot\vec{\phi} 
\end{array}
\right)\,.
\label{hamiltonian8}
\end{eqnarray}
$H$ is regarded as a effective Hamiltonian of the model. 
If background is flat ({\it i.e.},$B=1,B'=0,C=r$), 
the corresponding Hamiltonian easily reads 
\begin{eqnarray}
H_{\rm flat}=-\gamma^{\hat{6}}\gamma^{\hat{4}}\partial_4-\gamma^{\hat{6}}\gamma^{\hat{5}}\partial_5
+\gamma^{\hat{6}}m\vec{\tau}\cdot\vec{\phi}
\label{h_s}
\end{eqnarray}
where the partial derivatives $\partial_4,\partial_5$ in Cartesian coordinate are defined as
$\partial_4\pm i\partial_5=e^{\pm i\theta}(\partial_y\pm i\partial_\theta/y)$.
For obtaining the concise form (\ref{h_s}), 
we introduce and use the additional component of the gamma matrix $\gamma^{\hat{6}}:=I\otimes \sigma^3$.
One easily confirms that $H_{\rm flat}$ commutes with ``grandspin''
\begin{eqnarray}
K_3:=l_3+\frac{\gamma^{\hat{6}}}{2}+n\frac{\tau^3}{2},~~
l_3:= x_4p_5-x_5p_4
\end{eqnarray}
and also ``time-reversal operator''
\begin{eqnarray}
{\cal T}:=i\gamma^{\hat{4}}\otimes\tau^2 C
\end{eqnarray}
where $C$ is the charge conjugation operator. We emphasize that these operators also 
commute with the Hamiltonian in the curved space-time. As a consequence, eigenstates are specified by
the magnitude of the grandspin,{\it i.e.}, 
\begin{eqnarray}
&&K_3=0,\pm 1,\pm 2,\pm 3\cdots {\rm for~odd~} n \nonumber \\
&&K_3=\pm \frac{1}{2},\pm \frac{3}{2},\pm\frac{5}{2},\cdots {\rm for~even~} n\,.
\end{eqnarray}
Since the Hamiltonian is invariant under time reverse, the states of $\pm K_3$ are degenerate 
in energy. 

\begin{figure}
\includegraphics[height=7.0cm, width=9cm]{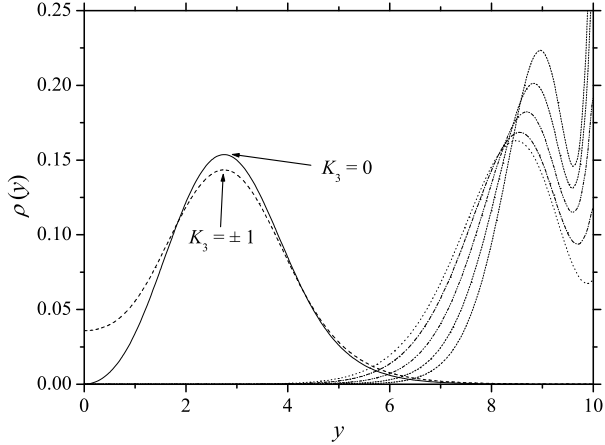}
\caption{\label{fig7} Fermion number density for the background skyrmion with $(m,n)=(1,3)$
where the coupling constant $m=0.76$. The states $K_3=0,\pm 1$ are localized on the 
brane core. Doubly degenerate $K_3=\pm 1$ states are more localized 
(they exhibit the non-zero value at the origin).}
\end{figure}

In order to treat the eigenproblem (\ref{eigenproblem8}), we employ the method which was 
originally proposed by Kahana-Ripka~\cite{kahana} for solving the Dirac equation with 
non-linear chiral background.
We construct the plane-wave basis in large circular box with radius $D$ 
as a set of eigenstates of the flat, unperturbed ($B=1,B'=0,C=r,f=\pi$) Hamiltonian {\it i.e.},
$H_0=-\gamma^{\hat{6}}\gamma^{\hat{4}}\partial_4-\gamma^{\hat{6}}\gamma^{\hat{5}}\partial_5
-\gamma^{\hat{6}}m\tau^3$. The solutions are 
\begin{eqnarray}
&&
\left(
\begin{array}{c}
u \\
v
\end{array}
\right)_{0,{\rm up}}
~=
M_{k_i}\left( 
\begin{array}{c}
\omega^{+}_{\epsilon_{k_i}}J_{p-1}(k_iy)e^{i(p-1)\theta} \\
\omega^{-}_{\epsilon_{k_i}}J_{p}(k_iy)e^{ip\theta}
\end{array}
\right)
\otimes
\left(
\begin{array}{c}
1 \\
0
\end{array}
\right) \nonumber \\
&&
\left(
\begin{array}{c}
u \\
v
\end{array}
\right)_{0,{\rm down}}
=
N_{l_i}\left( 
\begin{array}{c}
\omega^{-}_{\epsilon_{l_i}}J_{q}(l_iy)e^{iq\theta} \\
\omega^{+}_{\epsilon_{l_i}}J_{q+1}(l_iy)e^{i(q+1)\theta}
\end{array}
\right)
\otimes
\left(
\begin{array}{c}
0 \\
1
\end{array}
\right) \nonumber \\ 
\label{kahana_ripka}
\end{eqnarray}
with
\begin{eqnarray}
	M_{k_i}=\biggl[\frac{2\pi D^2\epsilon_{k_i}}{\epsilon_{k_i}+m}
	\Bigl(j_{p-1}(k_iD)\Bigr)^2\biggr]^{-1/2} \nonumber \\
	N_{l_i}=\biggl[\frac{2\pi D^2\epsilon_{l_i}}{\epsilon_{l_i}+m}
	\Bigl(j_{q+1}(l_iD)\Bigr)^2\biggr]^{-1/2} \nonumber 
\end{eqnarray}
and $\omega^{+}_{\epsilon_k<0},\omega^{-}_{\epsilon_k>0}=1, 
\omega^{-}_{\epsilon_k<0},\omega^{+}_{\epsilon_k>0}=-{\rm sgn}(\epsilon_k)k/(\epsilon_k+m)$.
The momenta $k_i,l_i~(i=1,\cdots,m_{\rm max})$ are discretized by the boundary conditions
\begin{eqnarray}
J_{p}(k_iD)=0,~~J_{q}(l_iD)=0
\end{eqnarray}
where $p:=K_3+\dfrac{1}{2}-\dfrac{n}{2}, q:=K_3-\dfrac{1}{2}+\dfrac{n}{2}$.
The orthogonality of the basis is then satisfied by  
\begin{eqnarray}
&&\int^D_0 dr r J_\nu(k_i r)J_\nu(k_j r)
=\int^D_0 dr r J_{\nu\pm 1}(k_i r)J_{\nu\pm 1}(k_j r)  \nonumber \\
&&=\delta_{ij}\frac{D^2}{2}  [J_{\nu\pm 1}(k_i D)]^2,~~
\nu:=K_3\pm\frac{1}{2}\mp\frac{n}{2}\,.
\label{orthogonality}
\end{eqnarray}
Expanding the eigenstates of Eq.(\ref{eigenproblem8}) in terms of the plane-wave basis, 
the eigenproblem reduces to the symmetric matrix diagonalization problem.  
A special care is taken on the estimation of the matrix element from the Hamiltonian 
(\ref{hamiltonian8}). In order to hold the Hermiticity of the matrix, 
the following differential rule is imposed
\begin{eqnarray}
\langle \psi |\overleftrightarrow{\partial_y}|\phi\rangle =\int dyd\theta \tilde{C}(y)
\frac{1}{2}\Bigl[\psi^{\dagger}\partial_y\phi-(\partial_y\psi^{\dagger}) \phi\Bigr]
\end{eqnarray}

\begin{figure}
\includegraphics[height=7.0cm, width=9cm]{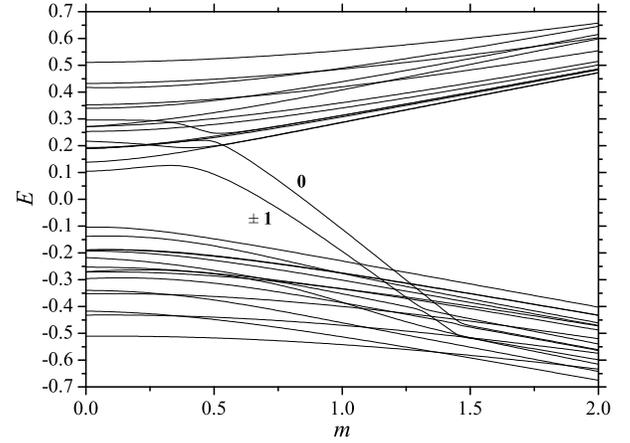}
\caption{\label{fig8} Spectral flow of the fermion energy of $(m,n)=(1,3)$, 
with some excited energy spectra. The zero crossing spectra have the quantum numbers $K_3=0,\pm 1$, 
which correspond to the localizing zero modes.}
\end{figure}

\begin{figure*}
\includegraphics[height=4.5cm, width=5.5cm]{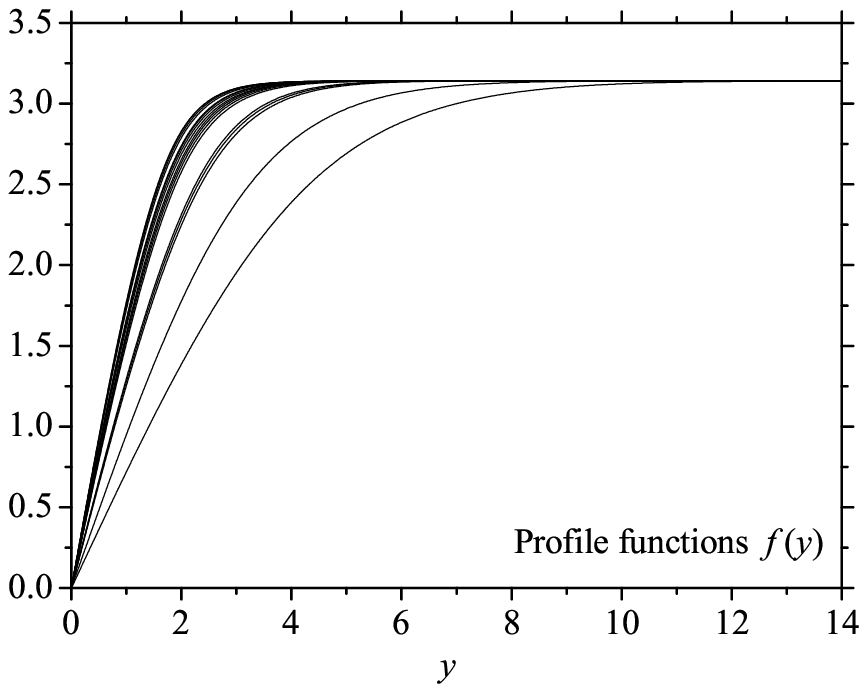}
\includegraphics[height=4.5cm, width=5.5cm]{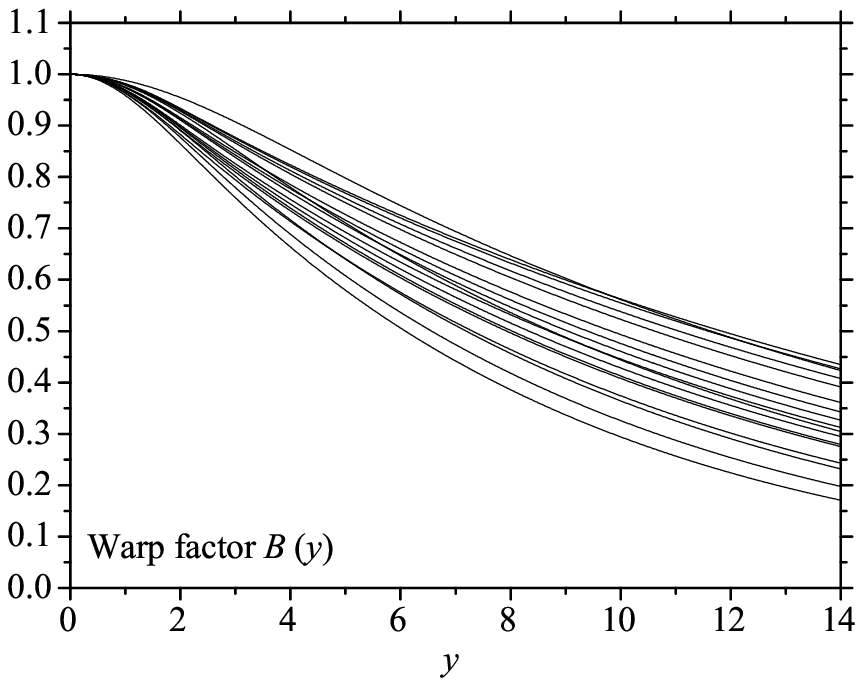}
\includegraphics[height=4.5cm, width=5.5cm]{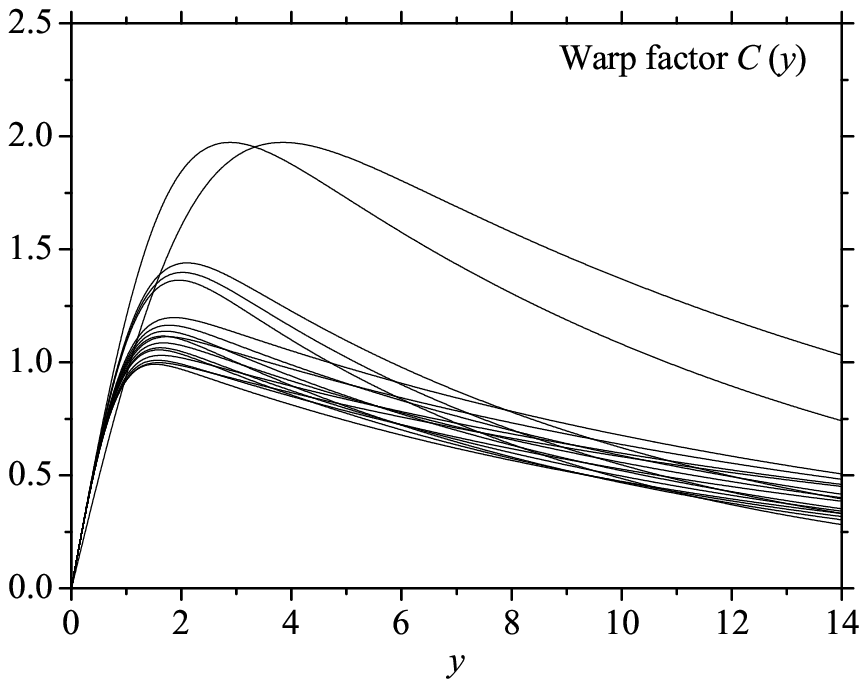}
\caption{\label{fig9} The brane solutions with large parameter space are plotted.}
\end{figure*}

Once the desired eigenfunctions are obtained, angular averaged fermion densities on the brane can 
be estimated as follows
\begin{eqnarray}
&& R(y) =N(y)\rho (y) \\
&&\rho (y):=\int d\theta[u^\dagger(y,\theta) u(y,\theta)+v^\dagger(y,\theta) v(y,\theta)]
\end{eqnarray}
where
\begin{eqnarray}
N(y)=\exp\biggl[-4\ln B(y)+\int^y \frac{dy'}{\tilde{C}(y')}\biggr]\,.
\end{eqnarray}

For a numerical convenience, we divide the extra space-time into two domains. 
We introduce an effective size of brane $D^{*}$ that the stress energy tensors are finite inside. 
Outside this domain, no brane exists and only the warping of the geometry is effective. 
We assume that the fermions we should observe in our 4D space-time are perfectly trapped 
inside and never leak outside. 
We set the size as $D^*\equiv D$ and tentatively choose the value $D=10$, which is larger than the 
distribution of the stress energy tensors and the topological charge (see Fig.\ref{fig2}).
Of course outside of the domain is not Minkowski space so that the fermions at the 
boundary feel the effects. The numerical results necessarily depend on the choice of $D$. 
This crude approximation works well if the effects of the geometry do not affect to the 
localized modes of the fermions. 

\begin{figure}
\includegraphics[height=7.0cm, width=9cm]{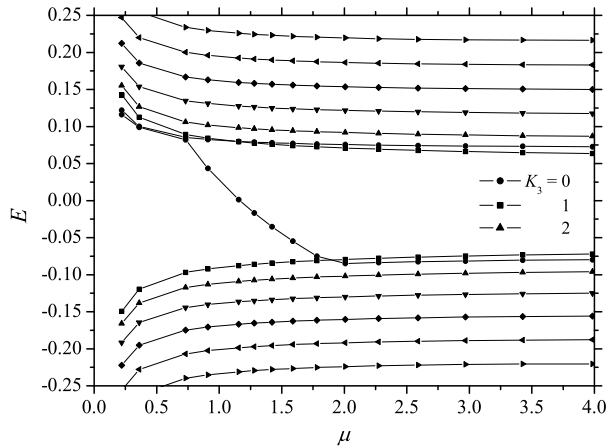}
\caption{\label{fig10} Spectral flow of the fermion energy of $(m,n)=(1,1)$ for 
which $m=1.0, \tilde{\Lambda}_b=-0.1$ and different values of $\alpha,\tilde{\mu}$.
Every dots indicates the fermion energy of the different background brane solutions. 
}
\end{figure}

For the numerical analysis except for the results of Figs.[\ref{fig9},\ref{fig10}], 
we use the brane solutions with following parameter set for the background fields
\begin{eqnarray}
&&(m,n)=(1,1):\nonumber \\
&&\hspace{1cm}\alpha=0.5,~\tilde{\Lambda}_b=-0.1,~\tilde{\mu}=0.220525915\,, \nonumber \\
&&\hspace{1cm}f'(0)=0.7292182131078184;\\
&&(m,n)=(1,3):\nonumber \\
&&\hspace{1cm}\alpha=0.11,~\tilde{\Lambda}_b=-0.1,~\tilde{\mu}=0.49588\,, \nonumber \\
&&\hspace{1cm}f^{(3)}(0)=0.1913003760462098.
\end{eqnarray}
The numerical results are shown in Figs.[\ref{fig2},\ref{fig8}].

Fig.\ref{fig3} shows the density $\rho(y)$ for the background skyrmion of $ (m,n)=(1,1)$.
We present a tower of the massive modes as well as the ground state. 
As one easily see that only the lowest mode is peaked on the brane while all other modes
escape from the core thus we cannot observe them. 

In Fig.\ref{fig4}, we plot the densities for some ranges of coupling constant $m$. 
As you see in the figure, only zero mode (or quasi zero modes) are localized and 
the others go far away from the brane.
In Fig.\ref{fig5}, we plot some crucial eigenvalues for varying $m$. The result 
clearly exhibits so called the level crossing picture~\cite{kahana}. 
Spectral flow is defined as the number of eigenvalues of Dirac Hamiltonian that 
cross zero from below minus the number of eigenvalues that cross 
zero from above for varying the properties of the background fields. The spectral 
flow exactly coincides with the topological charge and the zero modes 
are emerged when they cross the zero. 
In this case, for varying coupling constant $m$, one zero mode is observed. 

Libanov,Troitsky have discussed relation between the topological charge and the 
fermionic generation \cite{libanov01}. Four-dimensional fermions appear as zero-modes trapped in the 
core of the global vortex with winding number three. 
We shall investigate this speculation with our solution for $(m,n)=(1,3)$. 
The background brane solution is given in Fig.\ref{fig6}.
As is expected, we observe three localized solutions (Fig.{\ref{fig7}}) 
which can be regarded as the generations of the fermions.  
A main difference from the previous studies is that we use non-linear type of soliton solutions 
for constructing the branes. They have a richer information of the topology and 
the spectra exhibit doubly degenerate ground states and a higher excited state. On the other hand, 
the linear type of solitons like Abelian vortex have the spectra with double degeneracy only. 
Fermion generation of our universe comprises approximate doubly degenerate states 
and a very high energy state which
suggests the underlying topology that we employs.

More generally, we explore the fermion level behavior for various brane solutions with large parameter space.
In Fig.\ref{fig9} we present the parameter dependence of the solutions with $(m,n)=(1,1)$. 
By using these, we examine the fermion energies and show the results in Fig.\ref{fig10},
in which we present the level crossing for fixed value of $m,\tilde{\Lambda}_b$ 
(thus the each lines become the function of $\alpha,\tilde{\mu}$). 
The emergence of the zero mode is observed and it strongly depends on 
the property of the branes. 

By using the asymptotics of the warp factors (\ref{bound-inf1}), 
we easily speculate the behaviour $N(y)$ at zero and infinity~\cite{randjbar-daemi03}
\begin{eqnarray}
&&N(y\to 0)\to y \nonumber \\
&&N(y\to \infty)\to \frac{1}{\epsilon_1^4}\exp(4py+\frac{1}{p\epsilon_2}e^{py})\,.  
\end{eqnarray}
At the far from the core, 
the densities of the excited modes are more enhanced than the localized mode.

\section{Conclusion}
In this article, we have proposed new brane solutions in six dimensional space-time 
and have discussed about localization of the fermions on them. 
The brane have constructed by a baby-skyrmion, a generalization of a O(3) non-linear $\sigma$ model, 
and realizes regular warped compactification in 6 dim. anti-de Sitter space-time. 
The metric is non-factorizable with the warp factors either exponentially diverging 
or decaying in static solutions. But only exponentially decaying warp factors 
can allow gravity localization near the brane in the sense that the 6 dim. Planck 
mass takes a finite value. 
Such solutions can be obtained numerically if we impose suitable boundary conditions 
at the distance far from the brane and integrate in backward. 
Forward integration method is unstable as a unique set of boundary conditions 
at the origin does not produce an unique solution. 

Once the skyrmion branes were found, we studied the fermion localization on the
skyrmion branes. The Dirac equation in 6 dim. curved space-time was constructed in terms of 
introducing the vielbein and six dimensional generalization of the gamma matrices. 
To treat the massive modes directly, we employ the standard representation of the gamma 
matrices. In order to solve the eigenproblem, we introduce the plane wave basis in
a large circular box with radius $D$.  To study the speculation that the charge of the 
background skyrmions is identified as the generation of the fermions of our universe, 
we investigated the 
fermions with charge $Q=1,3$. This conjecture may be confirmed in terms of the fermion 
spectral flow. They exactly coincide with the topological charge and practically embody the 
localized modes. We found localized fermion modes corresponding to the
topological charge. For $(m,n)=(1,1)$, we observed the solutions with some parameter ranges, 
which means the existence of the massive modes as well as the massless one.
For $(m,n)=(1,3)$, three solutions localized on the brane are found. They are 
the lowest modes of doubly degenerate and a excited state, which approximately
agree with the experimental measurements of fermion masses. On the other hand, in the 
Abelian Higgs model one would expect the spectra include only double degeneracy.

In Ref.\cite{ringeval02}, the authors had an elaborate analysis for the massive modes in a 5 dim. 
anti-de Sitter space-time.
They set a domain that is defined by the effective potential, and then solved the Dirac equation with
inside and out. By imposing the boundary conditions at the surface, they obtained a number of 
massive modes.
In our case, the setup is more crude; for the numerical reason, we tentatively assumed that the 
fermions are confined within a hypothetical domain. 
If the size goes to infinity, there is no effect to the zero mode; 
the others will be delocalized. Therefore the massive modes
that we observe in Fig.\ref{fig4} might be an artifact of our numerical set up. On the other hand, 
for $(m,n)=(1,3)$ the lower three modes still localize on the vicinity at the origin; 
thus we successfully observe the massive localized mode.

\section{Acknowledgement}
We appreciate to Noriko Shiiki for giving us many useful advices and comments 
in the early stage of the research.

\end{document}